# Engineering Photon Avalanche Thresholds in Single Nanocrystals with Plasmonic Nanostructures

*Marcin Szalkowski*[1,2*], *Maciej Fryckowski*[1,2], *Małgorzata Misiak*[3], *Artur Bednarkiewicz*[3], *Dawid Piątkowski*[1,2], *Sebastian Maćkowski*[1]

[1]Nanophotonics Group, Institute of Physics, Department of Physics, Astronomy and Informatics, Nicolaus Copernicus University in Toruń, Toruń, Poland

[2]Institute of Advanced Studies, Nicolaus Copernicus University in Toruń, Toruń, Poland

[3]Institute of Low Temperature and Structure Research, Polish Academy of Sciences, Wrocław, Poland

* all correspondence should be addressed to *marszal@umk.pl*

KEYWORDS: Photon avalanche upconversion, super-resolution imaging, rare-earth-doped nanocrystals, plasmonics, metallic nanostructures

## ABSTRACT

Optical sub-diffraction imaging has become an essential tool in modern research. Recently, a new imaging strategy has been demonstrated that exploits the strongly nonlinear excitation–emission response of rare-earth–doped nanocrystals operating in the photon avalanche (PA) regime. This approach benefits from simple optical instrumentation, near-infrared excitation and emission, low cytotoxicity, and high photostability. However, its practical implementation is hindered by relatively high (10-100s of kW/cm2) excitation power densities required to initiate the PA process. We aim to mitigate this limitation through electromagnetic field enhancement provided by metallic nanostructures. The results of microscopic studies of individual photon-avalanching nanocrystals coupled to a silver island film demonstrate a greater than tenfold reduction in the excitation power density required to trigger the avalanche emission. These experimental results, corroborated by numerical modelling, which elucidates the role of plasmon-mediated electromagnetic field enhancement in modifying the photon avalanche dynamics, paves the way towards new generation of low excitation power probes for sub-diffraction imaging and sensing.

Photon avalanche (PA) emission is a process of energy up-conversion characterized by a threshold, highly nonlinear relation between excitation power density ($I_{exc}$) and emission intensity ($I_{em}$), as described by a power law: $I_{exc} = I_{em}^{S}$ [1]. The nonlinearity factor $S$, interpreted as the photon avalanche power-dependence slope, typically exceeds 10 (the nonlinearities of over 40 [2], or even 500 [3] have been reported) – numbers far exceeding those known in conventional multiphoton up-conversion processes. The recent demonstration of the photon avalanche in nanocrystals has opened new avenues for applications of this effect in thermometry [4,5], neuromorphic computing [6,7], biosensing [8,9], and force sensing[10,11]. Within this broad potential applicability palette, bioimaging has emerged as the most mature and widely explored area for avalanching nanocrystals (ANCs)[2,12–14]. In this context, ANCs retain all key advantages of well-established rare-earth-doped materials, including high photostability, low toxicity, operation in the near-infrared spectral range compatible with the biological tissue transparency window, absence of blinking, and extensive surface functionalization capacity[2,15,16].

The strong optical nonlinearity of PA emission enables sub-diffraction imaging [1,13], where the improvement of optical resolution results exclusively from the effective narrowing of the point spread function (PSF). According to the formula: $\delta = \lambda/(2 \cdot NA \cdot \sqrt{S}) = \delta_0/\sqrt{S}$, where $\lambda$ is the wavelength of the emitted light, $NA$ is the numerical aperture of the objective lens, $\delta_0$ and $\delta$ represent lateral imaging resolutions in linear and non-linear regimes, respectively[13]. Clearly, the stronger the non-linearity in a PA system, the higher the spatial resolution of the imaging.

One key advantage of the PA-based optical imaging is that it requires just a regular raster scanning confocal microscope with a single, diffraction limited photoexcitation beam. This straightforward operation stays in strong contrast to the most of commonly used super-resolution imaging concepts [17] developed recent years to mitigate the fundamental diffraction limit of the optical resolution. Indeed, stimulated emission depletion (STED)[18], stochastic optical reconstruction microscopy (STORM)[19], photoactivated localization microscopy (PALM)[20], structured illumination microscopy (SIM)[21], near-field methods[22], and their combinations[23], have led to breakthroughs in fluorescence-based imaging with resolutions down to tens of nm or even less[24]. However, most of these techniques require precisely tailored fluorescent labels and specialized lasers or optical beam shaping, or, on the other hand, sophisticated instrumentation complemented with efficient computing.

While promising, the PA-based optical imaging requires high pumping powers for reasonable photon counts (typical PA thresholds >100 kW/cm$^2$). On the other hand, PA materials exhibit slow luminescence rise times (in the range of tens of milliseconds). Both these factors are inherent to the PA mechanism, where the ground state absorption induced by non-resonant laser excitation is followed by efficient and resonant excited state absorption responsible for triggering the PA effect. Several ideas – focused mainly on material engineering – aimed at reducing the excitation power threshold have been recently tested, including optimization of dopant type and concentration[1,2,25,26] together with host-lattice engineering[1,3], enabling thresholds as low as tens of $kW/cm^2$. Our approach described in this work relies on external control of the PA threshold. Namely, we postulate and test experimentally this hypothesis that the PA threshold in single ACNs can be reduced by coupling with plasmon resonances in metallic nanostructures[27].

Plasmon excitations in noble metallic nanoparticles are known from their ability to locally enhance the electromagnetic field and thus to modify the optical properties of nearby emitters, including their emission rates[28–30]. This interaction depends on the distance between the emitter and the nanoparticle, with the net effect ranging from emission quenching for short (<5 nm) separations, to strong enhancement of emission for distances of 10-20 nm[27,31]. For larger separations, the impact of the plasmon excitations gradually weakens. Plasmon excitations have been commonly utilized to enhance signals in surface enhanced Raman scattering (SERS), as well as to enhance the absorption or emission rates of fluorophores[32], including for biomedical applications[33,34] or solar energy storage[35–37].

In this work, we evaluate the impact of the plasmon excitations in a silver islands film (SIF) on the optical properties of the single avalanching 8% $Tm^{3+}$ doped $NaYF_4$ nanocrystals (ANCs) deposited on the top of these metallic nanostructures. Importantly, separation between plasmonically active SIF and avalanching core ANCs was intentionally controlled by purposeful passivation of the PA NPs with an undoped $NaYF_4$ shell of 2 nm, 6 nm or 11 nm thickness deposited during the synthesis of the ANCs. We observed a significant shell thickness-dependent reduction of the avalanche threshold power of the ANCs interacting with SIF as compared to reference ANCs deposited on a bare glass substrate. The strongest, over 10-fold threshold power downshift was found for the ANCs without any shell. The presented results indicate huge potential of plasmon engineering of the avalanche emission in rare earths doped nanomaterials, offering feasible low power operation.

## Results

Transmission electron microscopy images of the synthesized ANCs with corresponding size distributions are presented in **Figure 1a-h**. The ANCs exhibit monodisperse, uniform shapes characteristic for hexagonal β-$NaYF_4$ nanocrystals. Crystal phase purity of all the ANCs is confirmed by XRD data (**Fig. S1**). The essential feature of the set of ANCs studied in this work was a gradual increase of the thickness of the passive shell (0 nm, (1.92 ± 0.54) nm, (6.15 ± 1.00) nm, and (10.63 ± 0.91) nm) covering the $Tm^{3+}$-doped core ((18.59 ± 0.68) nm). The shell not only protects the photoactive core of ANCs against energy trapping or quenching due to surface defects, ligands and surrounding medium [38], but also allows for precise, nanometer-scale control of the distance between the core of the ANCs and metallic nanoparticles.

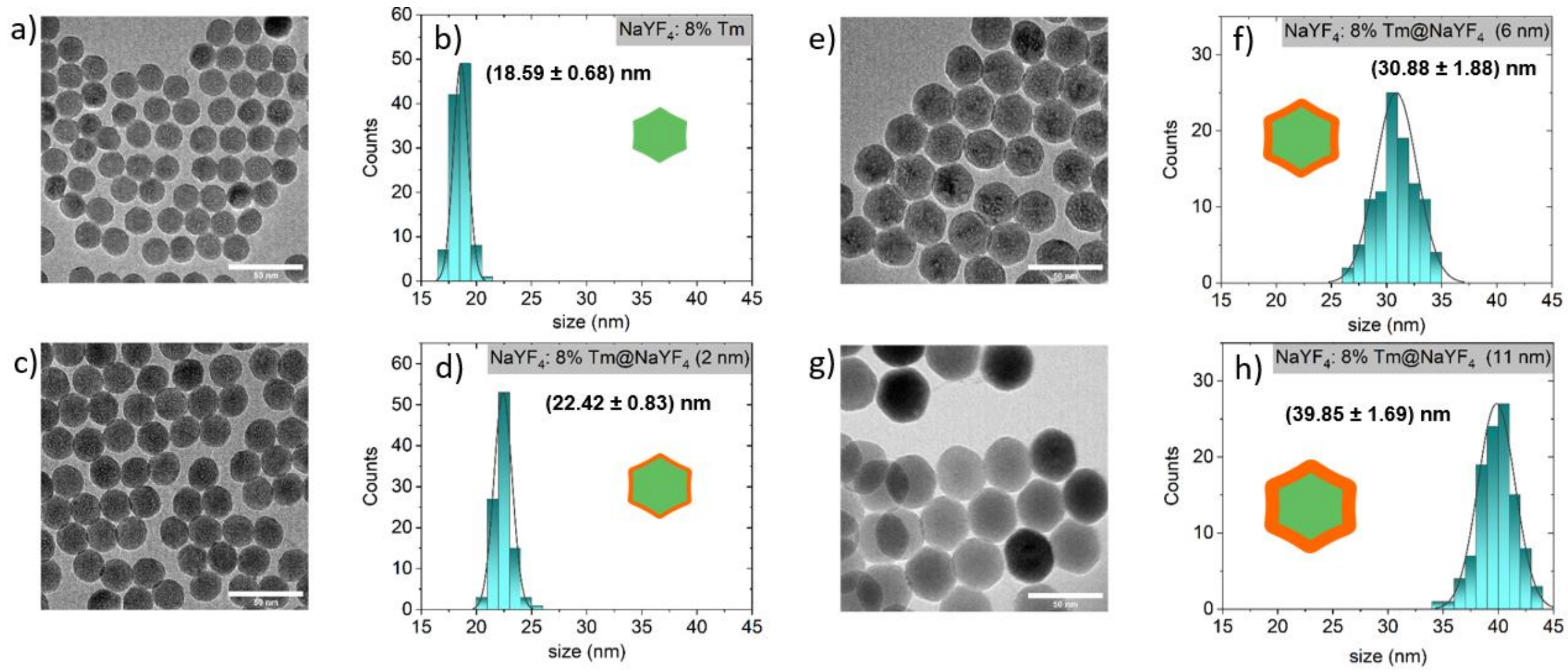

*Figure 1. TEM images (a, c, e, g) and corresponding size distributions (b, d, f, h) of ANCs of varying shell thickness.*

The energy diagram of the $Tm^{3+}$ ions exhibiting the PA emission [12,39,40], is presented in **Figure S2**. In the avalanching regime, the NC is pumped by a focused laser beam at 1059 nm (**Figure S3**, orange line), in resonance with the $^3F_4 \rightarrow {}^3F_{2,3}$ ($N_1 \rightarrow N_3$) excited state absorption (ESA), and simultaneously far from resonances to any ground state absorption (GSA). Initial population in the $^3F_4$ level is generated by residual side band absorption or established by thermal equilibrium. Avalanche emission (**Figure S3**, red line) with the main peak at 810 nm ($W_{R2}$), attributed to $^3H_4 \rightarrow {}^3H_6$ ($N_2 \rightarrow N_0$) transition, is powered by energy looping occurring by ($^3H_4$; $^3H_6$) $\rightarrow$ ($^3F_4$; $^3F_4$) cross-relaxation (CR) between the ions in the crystal matrix. This CR is responsible to continuously double the population of the $^3F_4$ metastable level after each ESA absorption event.

The transitions in the Tm-doped ANCs span over a broad spectral range, therefore, in order to facilitate efficient coupling between these transitions and plasmon resonances, we selected a silver island film (SIF) as metallic nanostructure [41,42]. Typically, SIF consists of irregular silver islands (30 nm – 250 nm) randomly deposited on a glass surface (**Fig. S4**). In addition to its rather simple fabrication, the SIF is a characterized with broad plasmon band, spanning from 400 nm to the near-infrared region, as shown in **Figure S3** (black). Although the wavelength of the excitation laser line is at 1059 nm, which is quite far away from the SIF's plasmonic resonance maximum, finite-difference time-domain (FDTD) numerical modelling (**Fig. S5-S7**) predicts strong plasmonic activity of the silver islands at these conditions, with the highest local field enhancements of over 60 for islands of c.a. 80 nm in diameter. Importantly, SIF exhibits a well-documented capability to strongly enhance the emission of broad variety of emitters[30,36,41–43].

A scheme of the sample, consisting of a SIF layer deposited on glass and covered with loosely dispersed single ANCs, is shown in **Figure S8**. To ensure exposure of the ANCs to the excitation beam while avoiding attenuation of the laser by the SIF layer, the sample was mounted on an oil-immersion objective in a configuration, where the ANCs were facing the objective through the immersion oil and protective coverslip. Because the refractive index of the immersion oil matches that of glass, such an oil-glass-oil multilayer arrangement should not introduce any distortions of the excitation beam. For reference, we also prepared an analogous reference sample with ANCs deposited directly on bare glass.

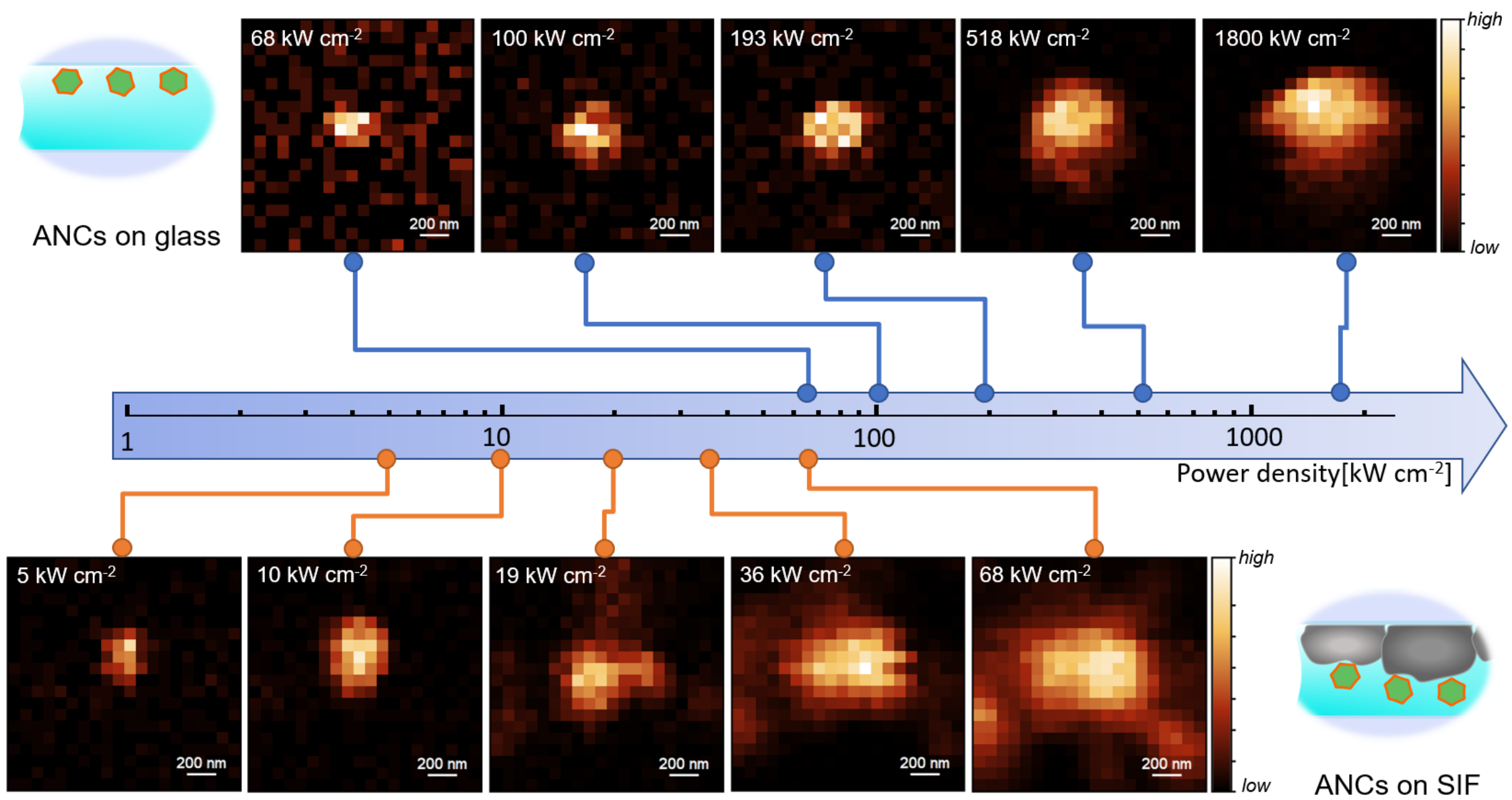

*Figure 2. Fluorescence images of single $NaYF_4$:$Tm^{3+}$ (8%) nanocrystals covered with a 2 nm thick passive shell and deposited on glass (upper row) or SIF (lower row) obtained at varying excitation power density.*

The central result of this experiment is displayed in **Fig. 2** and **Fig. S9**, where significant SIF-induced improvement of the luminescence imaging resolution upon gradually decreasing excitation power density is demonstrated. The images in upper row are luminescence raster scans of a single ANC with shell thickness of 2 nm deposited on glass substrate. Upon decreasing the excitation power from 1.8 MW $cm^{-2}$ to 68 kW $cm^{-2}$ the size of the optical image of the ANC is also reduced. This effect is accompanied with decrease of the emission intensity, i.e. for the lowest power used in this measurement the emission of the ANC was barely visible (signal-to-noise ratio, SNR, of 10). In a clear contrast, for the ANC placed on SIF (lower row) the same power density of 68 kW $cm^{-2}$ yields still very bright emission. Moreover, the optical signal of this ANC can still be clearly resolved for even lower excitation power of 5 kW $cm^{-2}$ (SNR = 44). This order-of-magnitude improvement observed for the ANC deposited on SIF confirms beneficial influence of plasmon excitations in SIF on the PA characteristics of single $Tm^{3+}$ ANCs. At the same time, for both ANCs (on SIF and on glass) we find gradual improvement of the imaging resolution with decreasing power, which corroborates previous predictions[13] and experimental observations[2,12,25,44].

In **Fig. 3** we show cross-sections of the emission intensity extracted for the ANCs shown in **Fig. 2** (**Fig. 3a** and **3b** for ANC on glass and on SIF, respectively). In both cases, reduction of the excitation power results in gradual narrowing of the size of the imaged ANC. This effect has been previously predicted [13] and demonstrated [12], and can be rationalized as a result of adjusting the pump power to fit into the most steep part of the *S*-shape, where *S* remains the largest. For too excessive excitation the saturation part of the *S*- shape dominates, therefore, owing to the low nonlinearity factor (S~2), resolution improvement is not observed. To quantify this effect, the cross-sections were approximated with Gaussian curves and the full widths at half maximum (FWHM) were extracted. For the ANC on glass FWHM decreased from 550 nm (for the 1.8 MW $cm^{-2}$ excitation) to 250 nm (for 68 kW $cm^{-2}$ excitation), corresponding to over a two-fold decrease in FWHM. In the case of the ANC on SIF, the respective value dropped from 480 nm (for the 68 kW $cm^{-2}$ excitation) to 230 nm (for the 5 kW $cm^{-2}$ excitation), yielding a comparable enhancement in spatial resolution, albeit much lower range of the excitation power. The result of such analysis carried out for a dozen of single ANCs with a given shell thickness deposited on glass or on SIF is presented in **Fig. 3c** (core ANCs), **Fig. 3d** (2 nm shell), **Fig. 3e** (6 nm shell), and **Fig. 3f** (11 nm shell), with error bars reflecting statistical variation. For all ANCs, irrespective of the shell thickness, a clear reduction of the FWHM (from c.a. 500 nm to 200 nm) with decreasing excitation power is visible, underlying the effect of power-dependent imaging resolution.

Concluding, while the imaging resolution improves with the excitation power, as expected for ANCs, in the case of ANCs deposited on SIF, this improvement occurs for substantially lower (5-10-fold) excitation powers. We associate this remarkable effect directly with the impact of the plasmon excitations in SIF on the optical properties of ANCs.

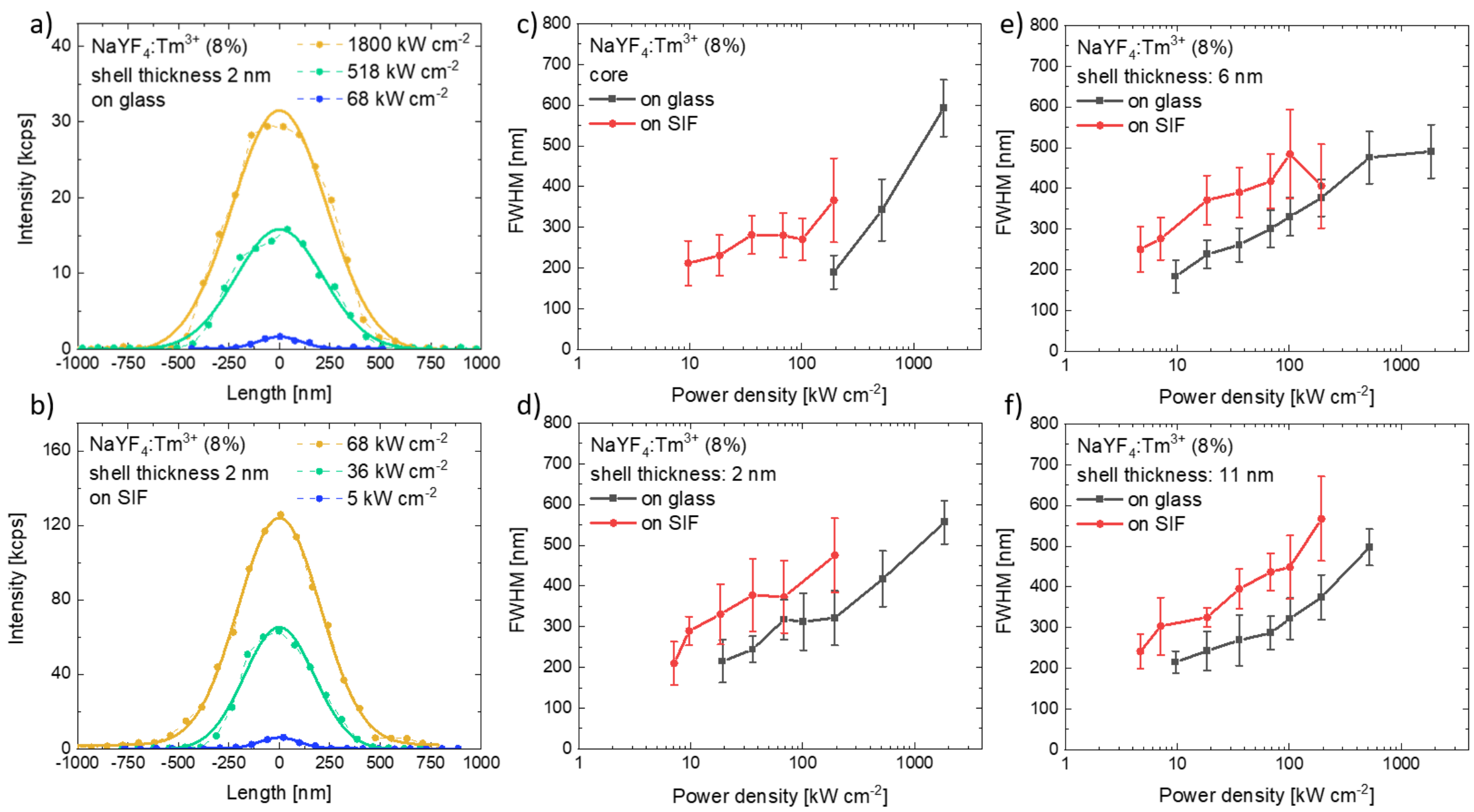

*Figure 3. Emission intensity cross-section for a single ANC placed on glass (a) and placed on SIF (b) for different excitation power densities are shown with dashed lines; corresponding Gaussian fits are presented with continuous lines. (c-f) panels show the power dependencies of average FWHM for the imaged individual ANCs of different shell thickness. Error bars represent the standard deviations (N~15).*

The reduction of the FWHM of spatial profiles with decreasing excitation power has been reported previously [1,12], and is one of the characteristic features of the ANCs, stemming from the strongly non-linear power – intensity dependence. The second one is power-dependent slowing down of the emission intensity rise times (**Fig. S10**), which is also present for both, reference

ANCs and ANCs placed on SIF, and reaches the values of few ms. The figure of merit for avalanching materials is however their *S*-shaped power dependence of the emission intensity [1,12]. In **Fig. 4a-d** we present the relation between the emission intensity and excitation power for single ANCs deposited on SIF (red) and on glass (black). Data obtained for ANCs synthesized with four shells of different thicknesses is displayed. For each configuration (substrate and shell thickness) a dozen of single ANCs was analysed with the distribution of experimentally obtained values marked as error bars. In all cases the *S*-shaped dependence appears, as expected for avalanching ANCs. Notably, for ANCs deposited on SIF a clear power downshift is observed as compared to the ANCs deposited on glass. This effect is most pronounced for ANCs with no shell (**Fig. 4a**) or with 2 nm shell (**Fig. 4b**). To quantify the influence of plasmon excitations in SIF on the emission intensity of single ANC, we multiply the results obtained for ANCs on glass to match the SIF data (shown by the dashed lines). The scaling factor was calculated to assure that the mean ratio ($R$) between the SIF and scaled glass datasets was the closest to unity. In this way, we may associate the multiplication factor with the strength of the interaction between ANCs and SIF. Its value should therefore depend on the shell thickness and indeed, such a dependence is observed (**Fig. 4e**). Interestingly, its shape seems analogous to the non-monotonic distance dependence of the fluorescence intensity, reported previously[2,12,13,44]. In **Fig. 4f** we plot a theoretical distance dependence of the emission enhancement for a single dipole interacting with a silver nanoparticle (Bharadwaj, *et al.*[31]) with the colours corresponding to emission quenching (blue), emission enhancement (red), and the region where there is no interaction present (grey). In order to understand the behaviour observed experimentally for single ANCs deposited on SIF it is important to realize that ANCs are not single dipole emitters. Instead, they contain hundreds of optically active ions within the relatively large (~20 nm) core, the size of which is comparable to the lengths characteristic for plasmonic interactions. This unique property of ANCs directly translates into a distribution of distances to the SIF for a single nanocrystal. Thus, each ion within the core experiences different interaction strength with plasmonic excitations in the SIF layer. Consequently, for a single nanocrystal placed in the vicinity of the SIF layer there are ions located very close to the SIF (possibly contributing to the emission quenching), as well as such that will interact with the SIF very weakly. This non-uniform character of plasmonic interactions in the studied system is included schematically as a colour coding in **Fig. 4a-d**, where the cores of the ANCs are shaded according to the range defined in **Fig. 4f**. In this case we also took into account the presence of the optically inactive shell.

In addition, high doping concentration of $Tm^{3+}$ ions which increases interionic interactions required to facilitate the avalanche emission, inhibits treating $Tm^{3+}$ ions embedded in ANCs as isolated emitters. Collective nature of the avalanche emission implies that affecting even the residual part of the ions might cause significant changes in overall avalanche characteristics of an individual NC[44]. Last but not least, while the passive shell plays a role of the separator between SIF and ANCs cores, the actual effective SIF-ANC distance is somewhat larger than the shell thickness due to the presence of the surface ligands stabilising the ANCs (c.a. 2 nm).

This semi-quantitative analysis indicates that both bare core and 2-nm shell ANCs should exhibit the strongest plasmonic effect, as the highest percentage of emitters in these two types of ANCs falls within the regime of strong enhancement (red colour). Nevertheless, for the former, the enhancement might be reduced by possible quenching occurring for the ions located at the closest proximity to the SIF surface (blue colour). In contrast, a thicker shell (**Fig. 4c-d**) results in much weaker stimulation of $Tm^{3+}$ ions by plasmon excitations, which translates into smaller reductions of the PA excitation power. This interpretation is consistent with the results of the luminescence decays measurements (**Fig. S11**). For all passive shell thicknesses shortening of the decays was observed, however, SIF-induced shortening was the most pronounced for bare core ANCs and ANCs covered with a 2 nm shell.

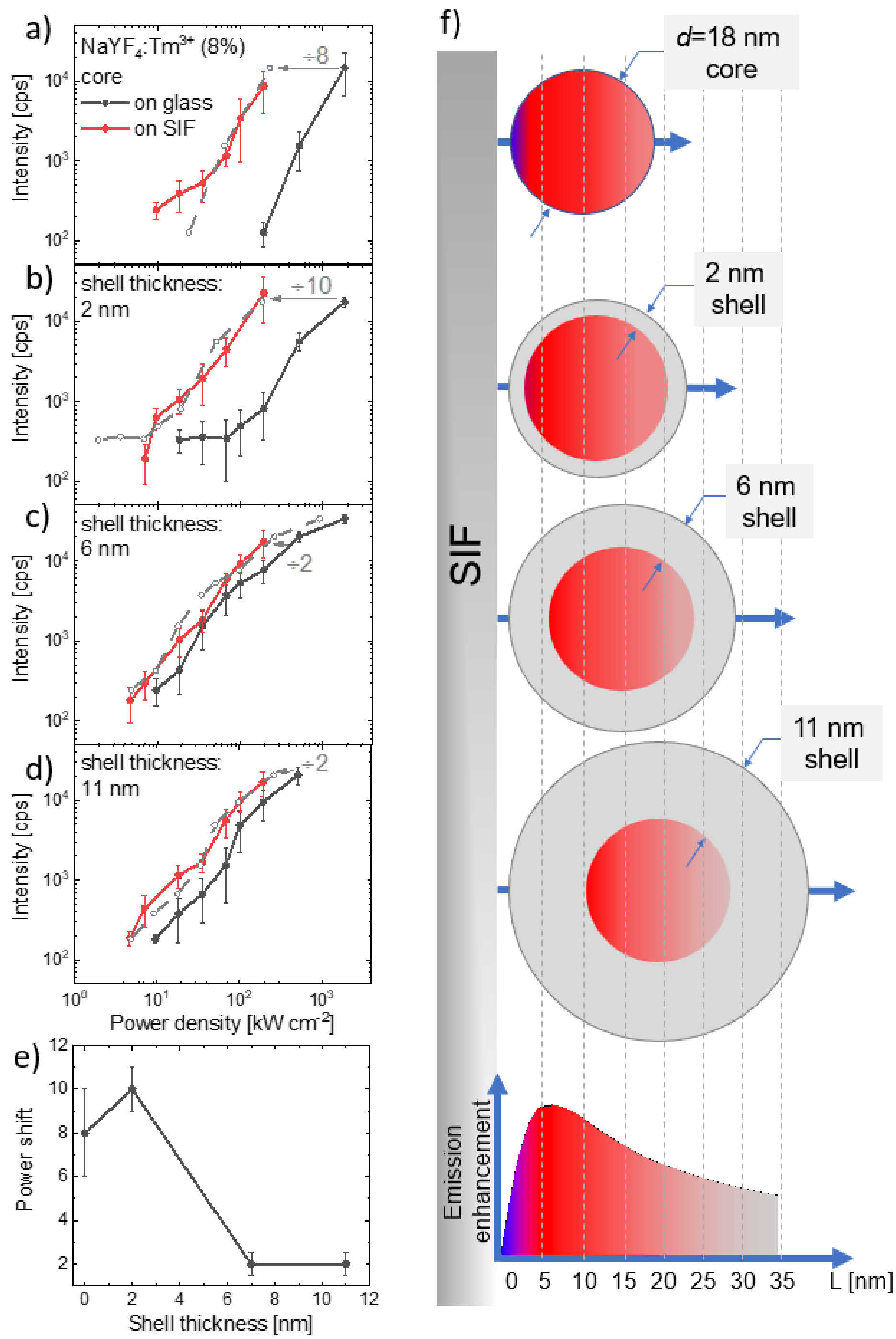

*Figure 4. a) – d) Averaged emission intensities of ANCs of varying shell thickness deposited on glass (black) or SIF (red). Gray dashed lines represent the results obtained on glass, but downshifted by the factor expressed on each image. e) Comparison of the power downshift factors in power dependences with varying shell thickness. f) Comparison of the theoretical distance*

*dependence of emission enhancement [31] with the size of the Tm-doped core of the ANC and its separation from plasmonically active substrate due to the presence of the passive shell and oleic acids. Error bars on (a-e) represent the standard derivations (N~15).*

Interactions with plasmon excitations influence – in the most general case – all the processes occurring in the emitter, and can be described using the following numerical model based on the set of differential rates equations (DREs)[12]. In this model we included three energy levels in $Tm^{3+}$ ($^3H_6$, $^3F_4$, $^3H_4$, marked as $N_0$, $N_1$ and $N_2$, respectively, as shown in **Fig. S2**), as population $^3F_{2,3}$ level is assumed to depopulate immediately to the lower laying $^3H_4$ state. The DREs set is as follows:

$$dN_0 = -(F + F^P)\sigma_{GSA}N_0 - (W_{CR} + W_{CR}^P)N_0N_2 + (W_{R1} + W_{NR1})N_1 + \big((1 - \beta_{21})(W_{R2} + W_{R2}^P)\big)N_2 + (Q_{11} + 2Q_{12}){N_1}^2 \quad \textbf{(Eq. 1)}$$

$$dN_1 = (F + F^P)\sigma_{GSA}N_0 - (F + F^P)\sigma_{ESA}N_1 + 2(W_{CR} + W_{CR}^P)N_0N_2 - (W_{R1} + W_{NR1})N_1 + (\beta_{21}(W_{R2} + W_{R2}^P) + W_{NR2})N_2 - (Q_{11} + Q_{12}){N_1}^2 \quad \textbf{(Eq. 2)}$$

$$dN_2 = (F + F^P)\sigma_{ESA}N_1 - (W_{CR} + W_{CR}^P)N_0N_2 - (W_{R2} + W_{R2}^P + W_{NR2})N_2 + Q_{12}{N_1}^2 \quad \textbf{(Eq. 3)}$$

$$N_0 + N_1 + N_2 = 1 \quad \textbf{(Eq. 4)}$$

The symbols are consistent with the **Fig. S2**. The ANC is exposed to $F=I_{exc}/h\nu$ photon flux. The model includes ground state ($\sigma_{GSA}$) and excited state ($\sigma_{ESA}$) absorption coefficients, as well as radiative ($W_{R1}$, $W_{R2}$) and nonradiative ($W_{NR1}$, $W_{NR2}$) relaxation rates, where lower indexes mark the initial level of the given transition. Moreover, $\beta_{21}$ is a branching ratio describing the percentage share of $N_2 \rightarrow N_1$ radiative transition in total $W_{R2}$ emission. $W_{CR}$ describes $(N_2, N_0) \rightarrow (N_1, N_1)$ cross-relaxation. Additionally, $Q_{12}$ and $Q_{11}$ represent the inverse CR processes [12]. All these parameters describe an isolated ANC. Exact values and definitions of the parameters used for the modelling are listed in SI (**Table S1**). To account for the interaction of ANCs with plasmons, we included the presence of plasmon-induced parameters marked with $P$ superscript. Importantly, the parameters emerging due to plasmon interactions are attributed to modification of the transition rates, except from $F^P$, which relates to local increase of the electromagnetic field density. A summary of the modelling results is presented in **Fig. 5** and **Fig. S9**, where power dependence curves obtained for isolated ANCs (f=0) and ANCs interacting with a metallic nanostructure are shown. Parameters describing plasmon-induced processes were expressed as a function of their respective counterparts from non-interacting reference system, e.g. $W_{R2}{}^P = f \times W_{R2}$, etc. Thus, for $f$=0, there is no plasmon-induced effect (the reference case of isolated ANC), while for $f$=10 a new interaction channel is introduced, being 10 times more efficient than in the reference. Such an approach, in which only one process-defining parameter is modified at a time, allows one to isolate how plasmon-induced changes influence the avalanche emission [44].

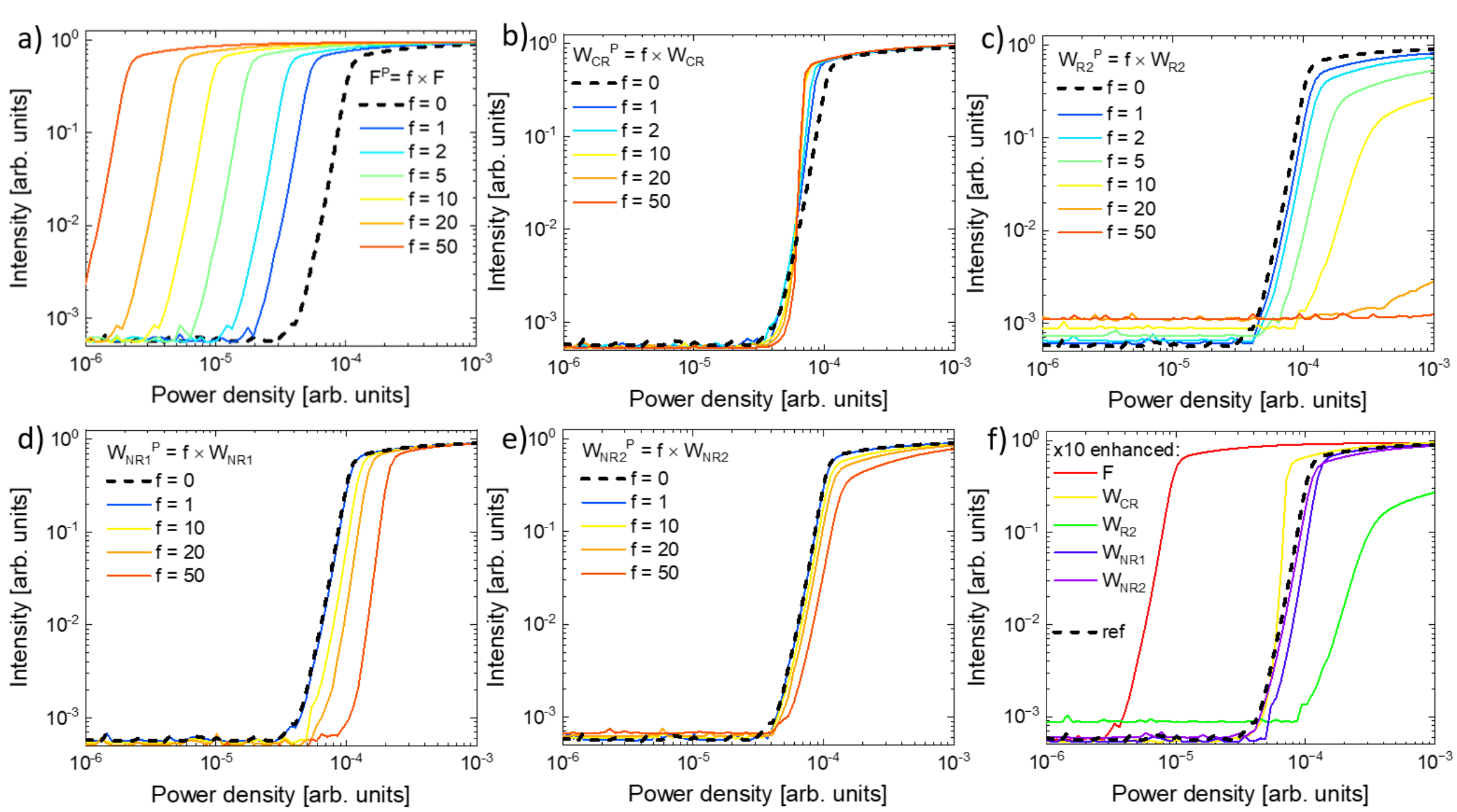

*Figure 5. Numerical DRE-based modelling of the photon avalanching system with selected processes enhanced due to interaction with plasmon excitation: $\sigma_{ESA}$, $W_{CR}$, $W_{R2}$, $W_{NR1}$ and $W_{NR2}$. Values of the plasmon-induced effective rates (marked with "P" superscript) were expressed as the value of the pristine values (for ANCs on glass) scaled by the f factor (f = 0, 1, 2, 5, 10, 20, 50). (f) Comparison of the results obtained for the f=10 for all considered rates.*

The presence of plasmon-enhanced electromagnetic field density ($F^P$) results in reduction of the PA threshold power, with the shape and slope barely affected (**Fig. 5a, Fig. S12**). Opposite effect was found for plasmon-induced modification of the $W_{CR}$ rate – with rising $W_{CR}^P$ value the power dependence curve does not experience severe up or down regulation of the threshold, but the nonlinearity is gradually rising (**Fig. 5b, Fig S12**). Analogous behaviour was observed for the enhancement of the processes related to radiative ($W_{R2}$) or nonradiative ($W_{NR1}$, $W_{NR2}$) depopulation of the excited states (**Fig. 5c-e, Fig. S12**). While low enhancements in interaction strength (up to 2, cyan curves) yield no significant changes, stronger plasmonic contribution increases the PA threshold. Within this group, the most destructive impact of the plasmonic interactions appears for enhanced radiative emission ($W_{R2}$ process), where enhancement higher than 2 results in significant reduction of the slope of the power dependence curve. Moreover, in this case emission intensity is affected in non-trivial way. While in pre-avalanche regime (below the threshold excitation power) emission intensity gradually rises with increased interaction strength *f*, for higher excitation powers opposite behaviour emerges. This might be non-intuitive that emission enhancement, which usually needs to be maximized to improve quantum efficiency of coupled emitters, here is the most parasitic one. However, this process hampers energy looping, which is the essence of avalanche emission, as it transfers ions from the emitting state directly to the ground state, without building up the population of the $N_1$ level. Moreover, the emission enhancement is simultaneously related to faster luminescence decays, which in turn hinders the CR processes. As a summary, **Fig. 5f** comprises and compares the impact of the plasmon-induced 10-fold change in the enhancements of the respective processes individually, on the overall performance of the avalanche luminescence. The strongest impact appears for enhanced $F$ and $W_{R2}$, (red and yellow curves, respectively), which are strongly shifting the PA threshold for lower ($F$) or higher ($W_{R2}$) excitation powers. In contrast, modification of other processes leads to rather negligible changes in power dependence. Our modelling served as guidance to understand how – and to which extend – each of those processes individually affect the PA behaviour. But it is important to emphasise, in realistic system, none of the energy transfer processes (R, NR, CR) can be addressed separately, like we did in our simulations, but rather they coexist and compete in a complex way. The only positive impact on the photon avalanche luminescence in terms of threshold power reduction was related to the plasmon-induced enhancement in local electromagnetic field density. This effect of light nano-condensing translates directly on the decrease of power density of the excitation beam required to observe avalanche emission. 10-fold reduction of the excitation power density is still much weaker than can be expected from simulations (**Fig. S5-S7**), where local field enhancement up to 60 was obtained for silver island. However, the studied structure is characterized by random positions of ANCs *vs* SIF, with the separation only roughly controlled by passive shell thickness and ligand length. Importantly, intensity of this enhanced field decays with rising distance (**Fig. S7**), what can be thus experienced only by part of the ions embedded in the host. To yield from stronger reduction of the excitation power, further optimization of the plasmon nanoparticles-ANCs nanostructures geometry would be required, to generate plasmonic hot spots and maximise their overlap with the photoactive avalanching cores.

## Conclusions

We presented optical microscopy studies of the interaction occurring between $Tm^{3+}$-doped ANCs and plasmonically active silver island film. Microscopic studies allowed to observe of ANCs clear features of the avalanche emission, including S-shaped power dependence and slowing down of emission intensity rise times for excitation powers close to the threshold. For these ANCs we demonstrated excitation power-dependent improvement of the imaging resolution. The same behaviour was found for ANCs deposited on SIF, however, for lower excitation powers than in reference structure. Threshold power reduction was tightly connected with the thickness of the passive shell covering the photoactive ANCs' cores. The most efficient interaction was found for the thickness of 2 nm, where 10-fold reduction of the excitation power threshold was obtained. Numerical modelling, based on DREs, allowed to assign the dominant part of observed effects to the plasmon-induced enhancement of the local electromagnetic field density. These results contribute to understanding the fundamentals of interaction of metallic plasmons with PA mechanism occurring in nanoscale particles, and demonstrate huge potential of engineering the properties of avalanching nanomaterials by interfacing them with plasmonically active nanostructures.

## Materials and methods

### *Synthesis and morphological characterization of the nanocrystals*

Commercially available reagents without any further purification were used for synthesis of nanocrystals. Yttrium oxide $Y_2O_3$ (99.99%), Thulium oxide $Tm_2O_3$ (99.99%) and ammonium fluoride $NH_4F$ (≥98.0%), oleic acid (90%) and 1-octadecene (90%) were purchased from Sigma Aldrich. Ethanol (EtOH, 96% pure p.a.), n-hexane (95%) and acetone (pure p.a.) were purchased from Chempur (Poland). Chloroform (pure p.a.) and sodium hydroxide NaOH (pure p.a.) were purchased from POCH S.A. (Poland). The lanthanide acetate precursors were prepared from lanthanide oxides. An appropriate amount of lanthanide oxides was dissolved in 50% aqueous acetic acid and placed in a Teflon-lined autoclave at 200°C for 3 hours. The solution was then transferred to a three-necked round-bottom flask, where excess water and acetic acid were evaporated, and then the precursor was dried on a hotplate at 140°C overnight. The synthesis was performed following literature protocols [45,46] with some modifications. 30 mL of oleic acid and 75 mL of 1-octadecene were added to a round-bottom flask containing 4.6 mmol of $Y^{3+}$ and 0.4 mmol of $Tm^{3+}$ acetate. The solution was heated to 145°C and stirred under vacuum for 30 minutes to form oleate salts and to remove water and oxygen. The temperature was decreased to ~80°C and the flask was placed under a nitrogen flow. Then, a solution containing 0.5 g of NaOH and 0.7410 g of $NH_4F$ in methanol was added. The solution was stirred in a closed flask under nitrogen at 70°C for 15 minutes, then the flask was opened and methanol was evaporated from the reaction mixture. The solution was then kept under vacuum for 15 minutes at 120°C. Afterwards, the temperature was increased to 305°C and the solution was kept further under nitrogen for one hour. After there, the mixture was cooled down to around 70°C and 16 mL of ethanol was added to precipitate the synthesized nanocrystals. The nanocrystals were then collected by centrifugation at 12 000 rcf for 10 minutes, washed with n-hexane and ethanol, centrifuged and dispersed in 10 mL of chloroform.

To synthesize core nanoparticles with a 2 nm shell, 40 mL of oleic acid and 40 mL of 1-octadecene were added to a round-bottom flask containing 5 mmol of yttrium trifluoroacetate and 5 mmol of sodium trifluoroacetate. The solution was heated to 120°C and stirred under vacuum for 30 minutes. The temperature was decreased to 80°C, and 9 mL of previously synthesized core nanoparticles were added. After the evaporation of chloroform from the solution, the temperature was increased to 300°C and the solution was stirred under a nitrogen atmosphere for one hour. The mixture was then cooled down, and 16 mL of ethanol was added to precipitate the nanocrystals. The nanocrystals were then collected by centrifugation at 12 000 rcf for 10 minutes, washed with n-hexane and ethanol, centrifuged and dispersed in 9 mL of chloroform. A similar protocol was used for the synthesis of 6 nm and 11 nm shells. The only changes were in the amount of nanocrystals added to the reaction mixture: 5 mL of core@ 2nm shell particles was used to synthesize core@ 6nm shell nanoparticles, and 9 mL of core@ 6nm shell nanoparticles was used to synthesize core@ 11nm shell nanoparticles. It should be underlined that all materials were prepared based on the same initial bare core ANCs, with the passive shell of given thickness produced step by step (0 nm → 2 nm → 6 nm → 11 nm), as described within this protocol.

The nanocrystals were structurally characterized by powder XRD measurements performed on a PANalytical X'Pert Pro diffractometer equipped with PIXcel ultrafast line detector, focusing mirror, and Soller slits for Cu Kα radiation. The morphology of the obtained samples was characterized using transmission electron microscope Philips CM20 SuperTwin at 160 kV.

*Fabrication of the silver islands film and preparation of the samples for imaging*

Silver island film was prepared with wet-chemistry method, as described previously[36,41,42,47]. The procedure is based on reducing silver nitrate ($AgNO_3$, purity >99% purchased from Sigma Aldrich) in the presence of NaOH (Acros Organics), ammonium hydroxide (25%, purchased from Acros Organics), and D-glucose (Fisher Scientific).

For sample preparation, a colloid of PA nanocrystals was diluted in ultrapure chloroform and spin-casted on a glass surface covered with SIF. Then, immersion oil was drop-casted on the top of this structure before covering with a bare glass coverslip and tightened to remove any excess oil from the space between the slides (**Figure S8**).

*Optical imaging*

Luminescence imaging was carried out with a custom-built confocal inverted optical microscope based on Ti2 Eclipse body (Nikon) equipped with a high *NA* oil immersion objective (CFI Apochromat TIRF 60XC Oil, *NA* = 1.49, Nikon). The sample was mounted on a piezoelectric stage (P-545, Physik Instrumente) for precise positioning and raster scanning during luminescence image acquisition. The emission of ANCs was excited using a pigtailed external cavity laser diode operating at 1060 nm (LPS-1060-FC, Thorlabs). The excitation beam was collimated, spatially filtered, and directed to the objective using a dichroic mirror (T926spxrxt, Chroma). Emitted light was collected by the same objective, filtered spatially by a 25 μm pinhole, and detected by a single photon counting module (SPCM, COUNT-100C, Laser Components). SPCM was synchronized with the piezoelectric stage and coupled to the pulse counter card (PCI-6236, National Instruments) for acquiring luminescence intensity or to the multiscaler card (MSA-300, Becker & Hickl) for collecting time decays of emission. For spectral analysis, emitted light was directed toward a monochromator (Shamrock 500i, Andor) equipped with a CCD camera (Newton, Andor).

## ASSOCIATED CONTENT

Supporting information is available in the *Supplementary material* file

AUTHOR INFORMATION

**Corresponding Author**

* all correspondence should be addressed to *marszal@umk.pl*

**Author Contributions**

MSz conceived and supervised the project, and designed the experiments. MM and MSz synthesized the materials and prepared the samples. MSz carried out optical measurements and processed the experimental results. MF and MSz carried out the numerical simulations. MSz, AB, DP and SM interpreted the data and coordinated the research activities. MSz wrote the original draft of the manuscript. MF, MM, AB, DP and SM reviewed and edited the manuscript. The manuscript was written through contributions of all authors. All authors have given approval to the final version of the manuscript.

ACKNOWLEDGMENT

AB and MM acknowledge the financial support from the National Science Center, Poland, within the 2021/43/B/ST5/01244 project. MSz and MF acknowledge the financial support from the National Science Center, Poland, within the 2024/55/D/ST11/02790 project.

ABBREVIATIONS

ANC, avalanche nanocrystal; CR; energy cross-relaxation; DRE, differential rate equation; ESA, excited state absorption; F, photon flux; FDTD, finite-difference time-domain; FWHM, full width at half maximum; GSA, ground state absorption; NR, non-radiative energy dissipation; PA, photon avalanche upconversion; PALM, photoactivated localization microscopy; PASSI, photon avalanche single beam super-resolution microscopy; PSF, point spread function; R, radiative emission; SIF, silver islands film; SIM, structured illumination microscopy; SNR, signal-to-noise ratio; STED, stimulated emission depletion; STORM, stochastic optical reconstruction microscopy.